\documentclass[fleqn,twoside]{article}
\usepackage{espcrc2,epsf,epsfig,psfrag,graphicx}
\usepackage[square,comma,sort&compress]{natbib}
\usepackage{graphicx}
\usepackage[figuresright]{rotating}
\newcommand{\beq}{\begin{equation}}
\newcommand{\eeq}{\end{equation}}
\newcommand{\bea}{\begin{eqnarray}}
\newcommand{\eea}{\end{eqnarray}}

\DeclareMathAlphabet{\mathsc}{OT1}{cmr}{m}{sc}
\newcommand{\CL} {C.L.}
\newcommand{\dof}{d.o.f.}
\newcommand{\eVq}{\rm{eV}^2}
\newcommand{\Sol}  {\mathsc{sol}}
\newcommand{\Atm}  {\mathsc{atm}}

\newcommand{\Dms}{\Delta m^2_\Sol}
\newcommand{\Dma}{\Delta m^2_\Atm}
\newcommand{\Dcq}{\Delta\chi^2}
\newcommand{\AHEP}{Instituto de F\'{\i}sica Corpuscular --
  C.S.I.C./Universitat de Val{\`e}ncia \\
  Edificio Institutos de Paterna, Apt 22085,
  E--46071 Val{\`e}ncia, Spain\\}
\newcommand{\snocc}{SNO$_\mathrm{CC}^\mathrm{rate}$ }
\newcommand{\snotot}{SNO$_\mathrm{CC,NC}^\mathrm{SP,DN}$ }
\def\lsim{\raise0.3ex\hbox{$\;<$\kern-0.75em\raise-1.1ex\hbox{$\sim\;$}}}
\def\gsim{\raise0.3ex\hbox{$\;>$\kern-0.75em\raise-1.1ex\hbox{$\sim\;$}}}
\def\e6{$E(6)$}
\def\10{$SO(10)$}
\def\21{$SU(2) \otimes U(1) $}

\def\422{$SU(4) \otimes SU(2) \otimes SU(2)$}
\def\321{$SU(3) \otimes SU(2) \otimes U(1)$}
\def\ne{\hbox{$\nu_e$ }}
\def\nm{\hbox{$\nu_\mu$ }}
\def\bne{\hbox{$\bar\nu_e$ }}

\def\nt{\hbox{$\nu_\tau$ }}

\def\ed{\end{document}}
\def \nbb {$\beta\beta_{0\nu}$ }

\newcommand{\AmS}{{\protect\the\textfont2
 A\kern-.1667em\lower.5ex\hbox{M}\kern-.125emS}}

\title{Standard and non-standard neutrino properties}

\author{J.~W.~F. Valle\address{\AHEP}
        \thanks{Work supported by Spanish grant PB98-0693,
by the European Commission RTN network HPRN-CT-2000-00148, by the
European Science Foundation network grant N.~86}}
       
\begin{document}

\begin{abstract}
  I review the interpretation of solar and atmospheric neutrino data
  in terms neutrino oscillations and describe some ways to account for
  the required neutrino masses and mixing angles from first
  principles, both within top-down and bottom-up approaches.  I also
  discuss non-oscillation phenomena such as \nbb which may probe the
  absolute scale of neutrino mass, and also reveal its Majorana
  nature.  I note that leptonic CP violation induced by ``Majorana''
  phases drop from oscillations but play a role in the leptogenesis
  scenario for the baryon asymmetry of the Universe.  Direct tests of
  leptonic CP violation in oscillation experiments, such as neutrino
  factories, will be a tough challenge, due to the hierarchical
  neutrino mass splittings and the smallness of $\theta_{13}$
  indicated by reactors.  The large solar mixing angle $\theta_{12}$
  offers a way to probe otherwise inaccessible features of supernova
  physics.  Finally, I note that in low-scale models of neutrino mass,
  one may probe \texttt{all} mixing angles, including the atmospheric
  $\theta_{23}$, at high energy accelerator experiments such as the
  LHC or NLC. A neat example is supersymmetry with bilinear breaking
  of R parity, where the LSP decay branching ratios are directly
  correlated to the neutrino mixing angles.  I also discuss
  non-oscillation solutions to the solar neutrino problem in terms of
  spin-flavor precession and non-standard neutrino interactions, which
  will be crucially tested at KamLAND.  \vspace{1pc}
\end{abstract}

\maketitle

\section{SOLAR NEUTRINOS}

Solar neutrinos have now been detected with the geochemical method
\cite{solgeo02} via the \ne + ${}^{37}$Cl $\to {}^{37}$Ar + $e^-$
reaction at Homestake, and via the \ne + ${}^{71}$Ga $\to ^{71}$Ge +
$e^-$ reaction at the Gallex, Sage and GNO experiments. Direct
detection with Cherenkov techniques using $\nu_e e$ scattering on
water at Super-K \cite{Smy}, and heavy water at SNO \cite{Hallin} have
given a robust confirmation that the number of solar neutrinos
detected in underground experiments is less than expected from
theories of energy generation in the sun \cite{Bahcall}. Especially
relevant is the sensitivity of the SNO experiment to the neutral
current (NC).

Altogether these experiments provide a solid evidence for solar
neutrino conversions and, as a result, imply that an extension of the
Standard Model of particle physics in the lepton sector is needed.

Although not yet unique, the most popular explanation of solar
neutrino experiments is provided by the neutrino oscillations
hypothesis.  Present data indicate that the mixing angle is
large~\cite{Maltoni:2002ni,Smirnov}, the best description being given
by the LMA MSW-type~\cite{Wolfenstein:1977ue} solution, already hinted
previously from the flat Super-K recoil electron
spectra~\cite{Gonzalez-Garcia:1999aj}.
The absence of a clear hint for day-night or seasonal variation
in the solar neutrino signal places important restrictions on neutrino
parameters.
%

%%%%%%%%%%%%%%%%%%%%%%%%%%%%%%%%%%%%%%%%%%%%%%%%%%%%%%%%%%%%%%%%%%%%%%
  
In ref.~\cite{Maltoni:2002ni} the solar neutrino data, including the
latest 1496--day data and the SNO NC result have been analysed in the
general framework of mixed active-sterile neutrino oscillations, where
the electron neutrino produced in the sun converts to a combination of
an active non-electron neutrino $\nu_x$ (a combination of $\nu_\mu$
and $\nu_\tau$) and a sterile neutrino $\nu_s \:$: $ \nu_e \to
\sqrt{1-\eta_s}\, \nu_x + \sqrt{\eta_s}\, \nu_s$.
The setting for such scenarios are four-neutrino mass
schemes~\cite{Peltoniemi:1992ss} which try to accommodate the solar
and atmospheric mass-splittings with the hint for short baseline
oscillations from LSND~\cite{LSND} indicating a large $\Delta m^2$.
The parameter $\eta_s$ with $0\le \eta_s \le 1$ describes the fraction
of sterile neutrinos taking part in the solar oscillations.

\begin{figure}[t] 
    \includegraphics[width=0.47\textwidth]{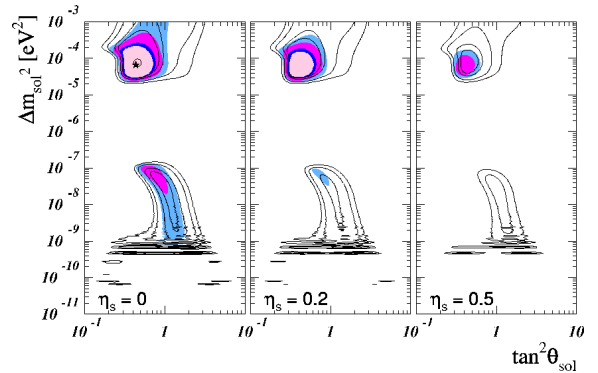}
\vspace*{-8mm}
      \caption{\label{fig:sol-osc-par}%
        Allowed $\tan^2\theta_\Sol$ and $\Dms$ regions for $\eta_s =
        0$ (active oscillations), $\eta_s = 0.2$ and $\eta_s = 0.5$.
        The lines and shaded regions correspond to the \snocc\ and
        \snotot analyses, respectively, as defined in
        Ref.~\cite{Maltoni:2002ni}.  The 90\%, 95\%, 99\% C.L. and
        3$\sigma$ contours are for 3 \dof.}
\vspace*{-5mm}
\end{figure}

In Fig.~\ref{fig:sol-osc-par} we display the regions of solar neutrino
oscillation parameters for 3 \dof\ with respect to the global
minimum, for the standard case of active oscillations, $\eta_s = 0$,
as well as for $\eta_s = 0.2$ and $\eta_s = 0.5$.
The first thing to notice is the impact of the SNO NC, spectral, and
day/night data in improving the determination of the oscillation
parameters: the shaded regions after their inclusion are much smaller
than the hollow regions delimited by the corresponding \snocc\ 
confidence contours. Especially important is the full \snotot\ 
information in closing the LMA region from above: values of $\Dms >
10^{-3}~\eVq$ appear only at $3\sigma$. Previously solar data on their
own could not close the LMA region, only the inclusion of reactor
data~\cite{CHOOZ} probed the upper part of the LMA
region~\cite{Gonzalez-Garcia:2000sq}. Furthermore, the complete
\snotot\ information is important for excluding \texttt{maximal} solar
mixing in the LMA region.  At $3\sigma$ we find the upper bound (1
\dof):
\begin{equation}
\label{t12bound}
    \rm{LMA}\,:\quad \tan^2\theta_\Sol \le 0.83 \,.
\end{equation}
In order to compare the allowed regions in Fig.~\ref{fig:sol-osc-par}
with others~\cite{Smirnov}, one must note that our \CL\ regions
correspond to the 3 \dof\ : $\tan^2\theta_\Sol$, $\Dms$ and $\eta_s$.
Therefore at a given \CL\ our regions are larger than the usual
regions for 2 \dof, because we also constrain the parameter $\eta_s$.
Our global best fit point occurs for active oscillations with 
\begin{equation}
\label{t12d12}
    \tan^2\theta_\Sol = 0.44\,, \quad
    \Dms = 6.6\times 10^{-5}~\eVq
\end{equation}

%%%%%%%%%%%%%%%%%%%%%%%%%%%%%%%%%%%%%%%%%%%%%%%%%%%%%%%%%%%%%%%%%%%%%%

\begin{figure}[t] 
    \includegraphics[width=0.47\textwidth]{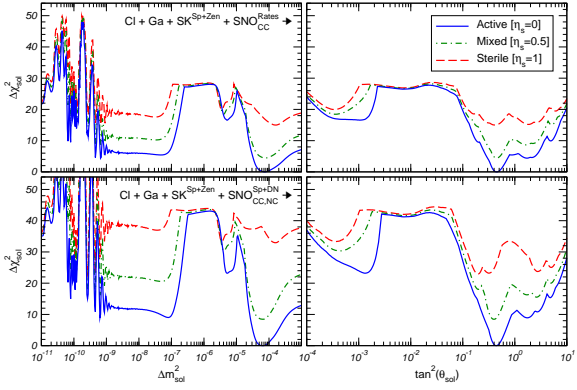}
\vspace*{-8mm}
  \caption{ \label{fig:chi-sol}%
    $\Delta\chi^2_\Sol$ as a function of $\Dms$ and
    $\tan^2\theta_\Sol$, for pure active ($\eta_s = 0$), pure sterile
    ($\eta_s = 1$) and mixed neutrino oscillations ($\eta_s = 0.5$).
    Upper and lower panels correspond to the \snocc\ and \snotot\ 
    samples defined in Ref.~\cite{Maltoni:2002ni}.}
\vspace*{-2mm}
\end{figure}

\begin{figure}[t] 
  \includegraphics[width=0.96 \linewidth]{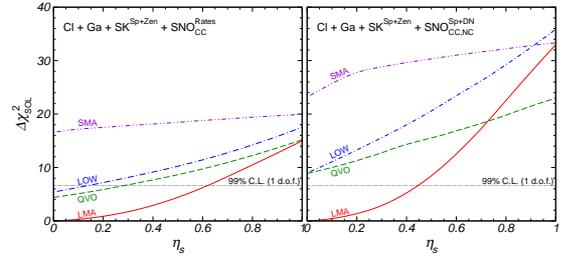}
\vspace*{-4mm}
  \caption{ \label{fig:sol02.etas}%
    $\Delta\chi^2_\Sol$ displayed as a function of $\eta_s$ with
    respect to favored active LMA solution, for the \snocc\ (left
    panel) and the \snotot\ (right panel) analysis, as defined in
    Ref.~\cite{Maltoni:2002ni}.}  
\vspace*{-2mm}
\end{figure}

A concise way to illustrate the above results is displayed in
Fig.~\ref{fig:chi-sol}. We give the profiles of $\Delta\chi^2_\Sol$ as
a function of $\Dms$ (left) as well as $\tan^2\theta_\Sol$ (right), by
minimizing with respect to the undisplayed oscillation parameters, for
the fixed values of $\eta_s=0$, $0.5$, $1$.
By comparing top and bottom panels one can clearly see the impact of
the full \snotot\ sample in leading to the relative worsening of all
non-LMA solutions with respect to the preferred active LMA solution.
One sees also how the preferred LMA status survives in the presence of
a small sterile admixture characterized by $\eta_s$ (also seen in
Figs.~\ref{fig:sol-osc-par} and \ref{fig:sol02.etas}). Increasing
$\eta_s$ leads to a deterioration of all oscillation solutions.

It is also instructive to display the profile of $\Delta\chi^2_\Sol$
as a function of $0 \leq \eta_s \leq 1$, irrespective of the detailed
values of the solar neutrino oscillation parameters $\Dms$ and
$\theta_\Sol$, as shown in Fig.~\ref{fig:sol02.etas}.
One can see that there is a crossing between the LMA and
quasi--vacuum--oscillations (QVO) solutions. This implies that the
best pure--sterile description lies in the QVO regime. However, in the
global analysis pure sterile oscillations with $\eta_s=1$ are highly
disfavored. We find a $\chi^2$-difference between pure active and
sterile of $\Delta\chi^2_\mathrm{s-a} = 32.9$ if we restrict to the
LMA solution, or $\Delta\chi^2_\mathrm{s-a} = 22.9$ if we allow also
for QVO. For 3 \dof\ the $\Delta\chi^2_\mathrm{s-a} = 22.9$ implies
that pure sterile oscillations are ruled out at 99.996\% \CL\ compared
to the active case. From the figure we obtain the bound
\begin{equation}
    \label{eq:etasSol}
    \rm{solar \: data:}\quad \eta_s \leq 0.45 \,.
\end{equation}
at 99\% \CL\ for 1 \dof. A complete table of best fit values of $\Dms$
and $\theta_\Sol$ with the corresponding $\chi^2_\Sol$ and GOF values
for pure active, pure sterile, and mixed neutrino oscillations is
given in \cite{Maltoni:2002ni}, both for the \snocc\ ($48-2$ \dof) and
the \snotot\ analysis ($81-2$ \dof).

%%%%%%%%%%%%%%%%%%%%%%%%%%%%%%%%%%%%%%%%%%%%%%%%%%%%%%%%%%%%%%%%%%%%%%

\section{ATMOSPHERIC NEUTRINOS}

%%%%%%%%%%%%%%%%%%%%%%%%%%%%%%%%%%%%%%%%%%%%%%%%%%%%%%%%%%%%%%%%%%%%%%

Here I summarize the analysis of atmospheric data given in
Ref.~\cite{Maltoni:2002ni}, in a generalized oscillation scheme
in which a light sterile neutrino takes part in the oscillations,
under the approximation $\Dms\ll\Dma$.  In order to comply with the
strong constraints from reactor experiments~\cite{CHOOZ} we completely
decouple the electron neutrino from atmospheric oscillations.
In contrast with the case of solar oscillations, the constraints on
the $\nu_\mu$--content in atmospheric oscillations are not so
stringent.  Thus the description of atmospheric neutrino oscillations
in this general framework requires two parameters besides the standard
2-neutrino oscillation parameters $\theta_\Atm$ and $\Dma$.  We will
use the parameters $d_\mu$ and $d_s$ already introduced in
Ref.~\cite{Maltoni:2001bc}, and defined in such a way that $1-d_\mu$
($1-d_s$) corresponds to the fraction of $\nu_\mu$ ($\nu_s$)
participating in oscillations with $\Dma$. Hence, pure active
atmospheric oscillations with $\Dma$ are recovered when $d_\mu=0$ and
$d_s=1$. In four-neutrino models there is a mass scheme-dependent
relationship between $d_s$ and the solar parameter $\eta_s$. For
details see Ref.~\cite{Maltoni:2001bc}.

To get a feeling on the physical meaning of these two parameters, note
that for $d_\mu=0$ we obtain that the $\nu_\mu$ oscillates with $\Dma$
to a linear combination of $\nu_\tau$ and $\nu_s$ given as
$ \nu_\mu \to \sqrt{d_s} \,\nu_\tau + \sqrt{1-d_s} \,\nu_s \,.$

\begin{figure}[t] 
  \includegraphics[width=0.46\textwidth,height=5cm]{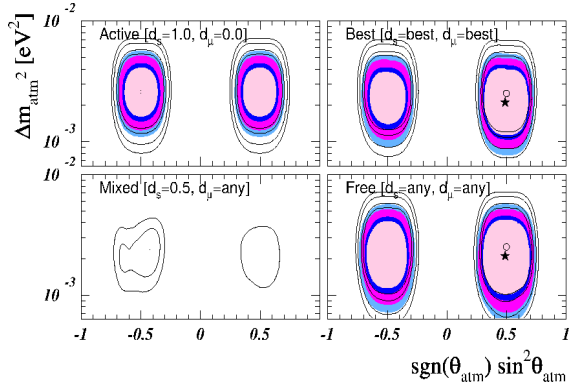}
\vspace*{-4mm}
  \caption{ \label{fig:atm-osc-par}%
    Allowed regions of $\sin^2\theta_\Atm$ and $\Dma$ at 90\%, 95\%,
    99\% and 3$\sigma$ for 4 \dof\ and different assumptions on the
    parameters $d_s$ and $d_\mu$, from~\cite{Maltoni:2002ni}. The
    lines (shaded regions) correspond to 1289 (1489) days of Super-K
    data. }  \vspace*{-2mm}
\end{figure}

\begin{figure}[t] 
  \includegraphics[width=0.46\textwidth,height=4cm]{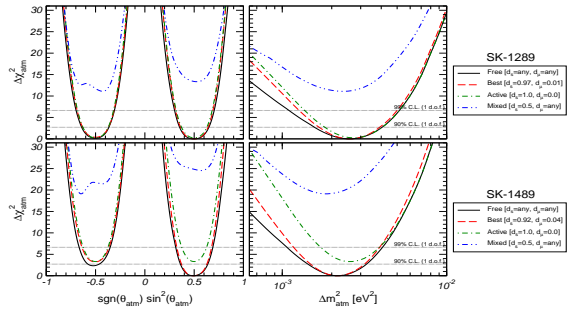}
  \vspace*{-5mm}
\caption{\label{fig:chi-atm}%
    $\Delta\chi^2_\Atm$ as a function of $\sin^2\theta_\Atm$ (left)
    and $\Dma$ (right), using 1289 (upper) and 1489 (lower) days of
    Super-K data~\cite{Maltoni:2002ni}.}  \vspace*{-2mm}
\vspace*{-2mm}
\end{figure}

Our global best fit point occurs at 
\begin{equation}
    \sin^2\theta_\Atm = 0.49 \,,\quad \Dma = 2.1 \times
    10^{-3}~\eVq 
\end{equation}
and has $d_s=0.92,\: d_\mu=0.04$. We see that atmospheric data prefers
a small sterile neutrino admixture. However, this effect is not
statistically significant, since the pure active case ($d_s=1,
d_\mu=0$) also gives an excellent fit: the difference in $\chi^2$ with
respect to the best fit point is only $\Dcq_\mathrm{act-best} = 3.3$.
For the pure active best fit point we obtain,
\begin{equation}
\label{t23d23}
    \sin^2\theta_\Atm = 0.5 \,,\: \Dma = 2.5 \times
    10^{-3}~\eVq \: 
\end{equation}
with the 3$\sigma$ ranges (1 \dof)
\begin{eqnarray} 
    0.3 \le \sin^2\theta_\Atm \le 0.7 \\ 
    1.2 \times 10^{-3}~\eVq \le \Dma  \le 4.8 \times
    10^{-3}~\eVq \,. 
\end{eqnarray}
The determination of the parameters $\theta_\Atm$ and $\Dma$ is
summarized in Figs.~\ref{fig:atm-osc-par} and \ref{fig:chi-atm}.  
Note that Fig.~\ref{fig:chi-atm} considers several cases: arbitrary
$d_s$ and $d_\mu$, best--fit $d_s$ and $d_\mu$, and pure active and
mixed active--sterile neutrino oscillations, as indicated.

%%%%%%%%%%%%%%%%%%%%%%%%%%%%%%%%%%%%%%%%%%%%%%%%%%%%%%%%%%%%%%%%%%%%%%

At a given \CL\ we cut the $\chi^2_\Atm$ at a $\Dcq$ determined by 4
\dof\ to obtain 4-dimensional volumes in the parameter space of
($\theta_\Atm, \Dma, d_\mu,d_s$).  In the upper panels we show
sections of these volumes at values of $d_s=1$ and $d_\mu=0$
corresponding to the pure active case (left) and the best fit point
(right). Again we observe that moving from pure active to the best fit
does not change the fit significantly.  In the lower right panel we
project away both $d_\mu$ and $d_s$, whereas in the lower left panel
we fix $d_s=0.5$ and eliminate only $d_\mu$.
Comparing the regions resulting from 1489 days Super-K data (shaded
regions) with the one from the 1289 days Super-K sample (hollow
regions) we note that the new data leads to a slightly better
determination of $\theta_\Atm$ and $\Dma$. However, more importantly,
from the lower left panel we see, that the new data shows a much
stronger rejection against a sterile admixture: for $d_s=0.5$ no
allowed region appears at 3$\sigma$ for 4 \dof.

\section{NEUTRINO PARAMETERS}

The basic structure of the neutrino sector required to account for
present solar and atmospheric data has been developed in the early
eighties, motivated mainly by
theory~\cite{Weinberg:uk,seesaw,seesaw2,seesaw3}.  In a gauge theory
of the weak interaction the \texttt{simplest} lepton mixing matrix is
characterized by 3 angles and 3 CP violating
phases~\cite{Schechter:1980gr}:
\begin{itemize}
\item the solar angle $\theta_{12}$
\item the atmospheric angle $\theta_{23}$
\item the reactor angle $\theta_{13}$
\item one Kobayashi-Maskawa-like CP phase 
\item 2 extra (Majorana-type) CP phases
\end{itemize}
The structure of leptonic weak interactions is more complex in
theories containing \21 singlet leptons~\cite{Schechter:1980gr},
especially if, for some symmetry reason, they happen to be light.  In
such case the charged current mixing matrix is rectangular and the
neutral current is non-trivial, with yet new angles and phases present
(see~\cite{Schechter:1980gr} for a detailed discussion and
parametrization).

%%%%%%%%%%%%%%%%%%%%%%%%%%%%%%%%%%%%%%%%%%%%%%%%%%%%%%%%%%%%%%%%%%%%%%

As seen above~\cite{Bilenky:1998dt}, current solar and atmospheric
data fit very well with oscillations among the three active neutrinos,
provided that $\theta_{12}$ is large as seen in Eq.~(\ref{t12d12}),
but non-maximal, given in Eq.~(\ref{t12bound}), while $\theta_{23}$
must be nearly maximal, from Eq.~(\ref{t23d23}). As mentioned,
$\theta_{13}$ must be rather small~\cite{Gonzalez-Garcia:2000sq}.
Note from Eqs.~(\ref{t12d12}) and~(\ref{t23d23}) that $\Dma \gg \Dms$.
Depending on the sign of $\Delta m_{32}^2$ there are two possible
neutrino mass schemes: normal and inverse-hierarchical neutrino
masses.

If solar and atmospheric data are combined with short baseline data
including the LSND hint, then one needs, in the framework of the
oscillation hypothesis, the existence of a light sterile
neutrino~\cite{Peltoniemi:1992ss} taking part in the
oscillations~\cite{Maltoni:2001bc}. 

With the latest data one finds that, even though 4-neutrino models can
not be ruled out as such, the resulting global fit of all current
oscillation data is extremely poor~\cite{Maltoni:2002xd}.

%%%%%%%%%%%%%%%%%%%%%%%%%%%%%%%%%%%%%%%%%%%%%%%%%%%%%%%%%%%%%%%%%%%%%%

\begin{figure}[t] 
\includegraphics[width=0.23\textwidth]{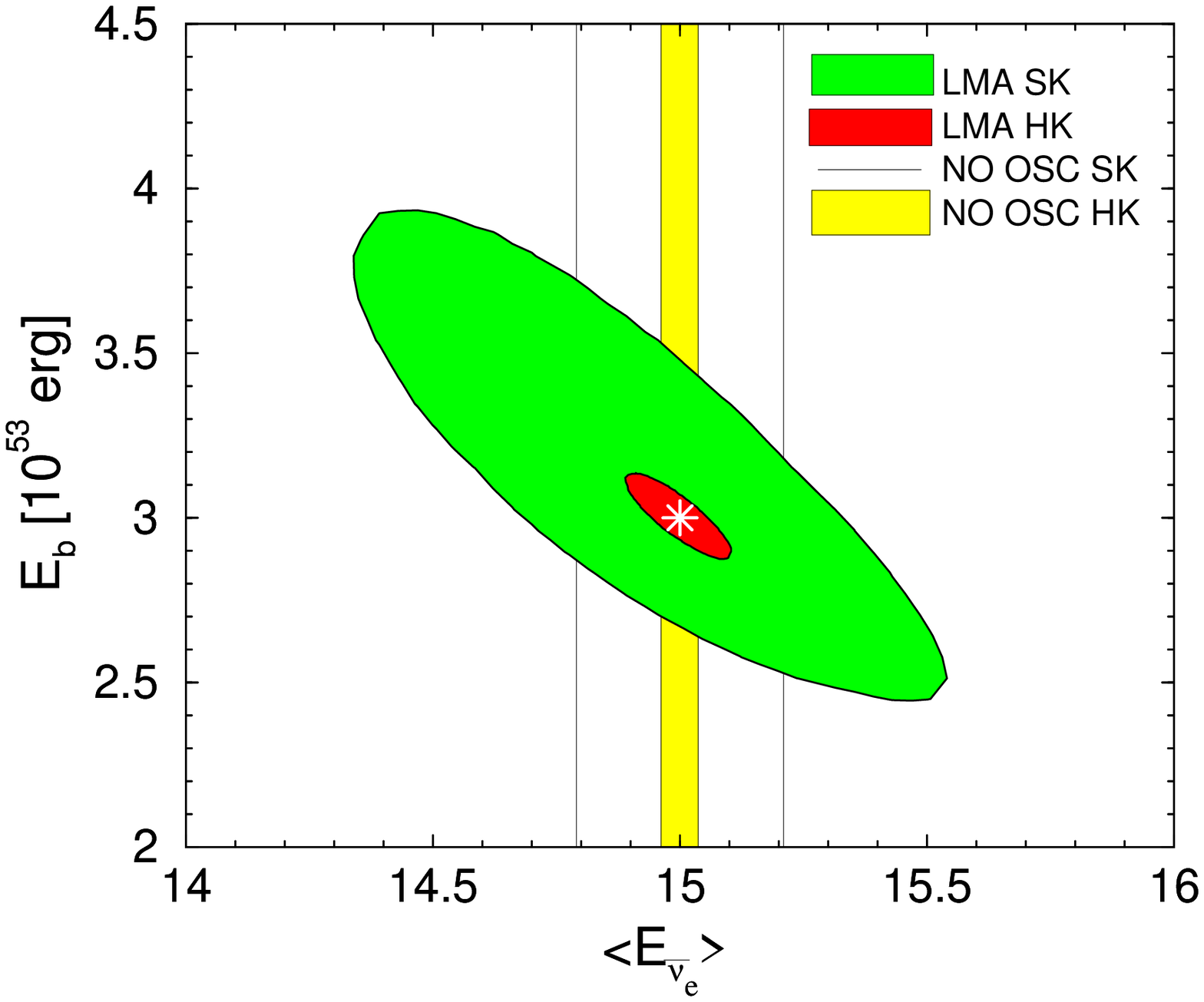}
\includegraphics[width=0.23\textwidth]{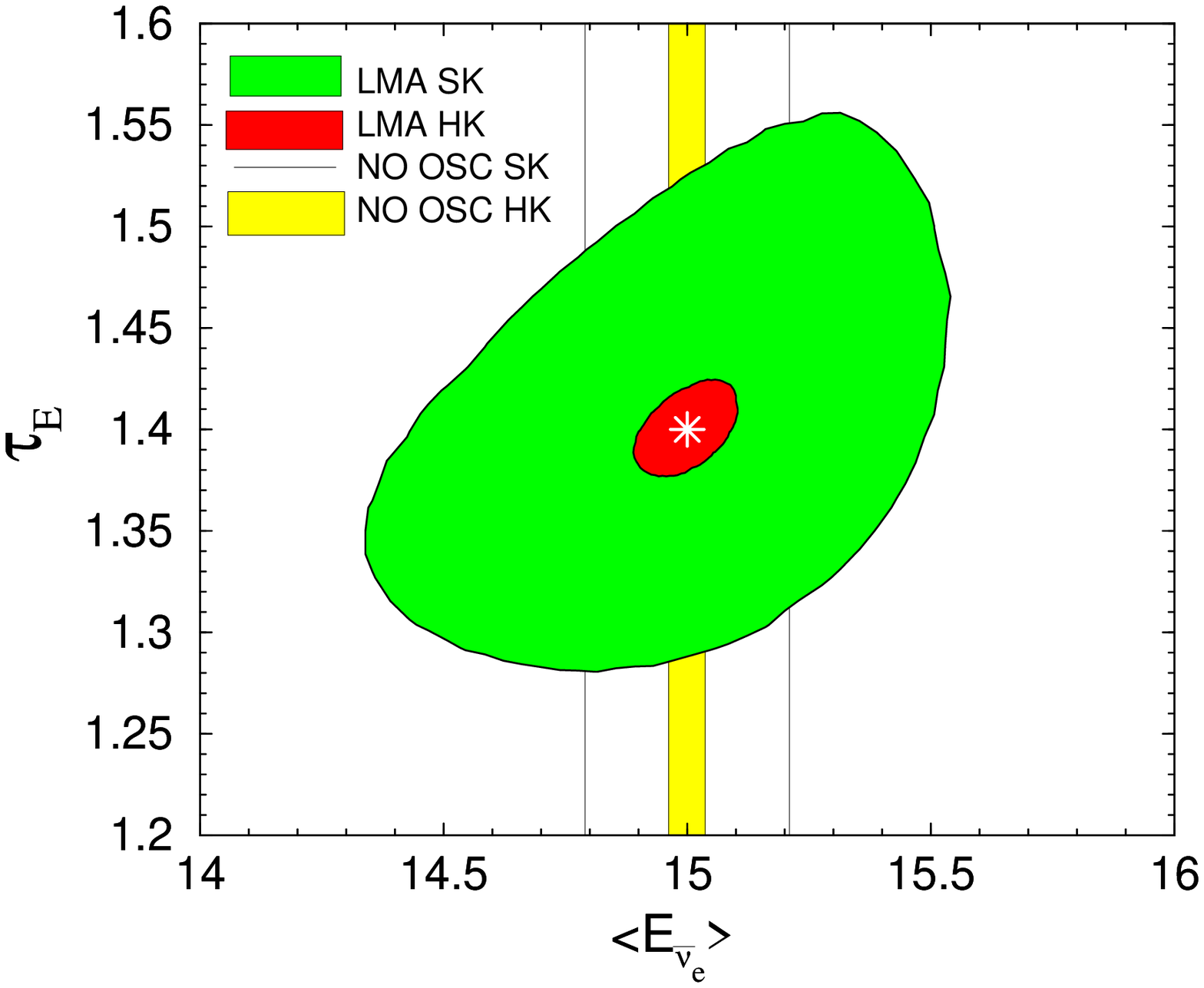}
\includegraphics[width=0.23\textwidth]{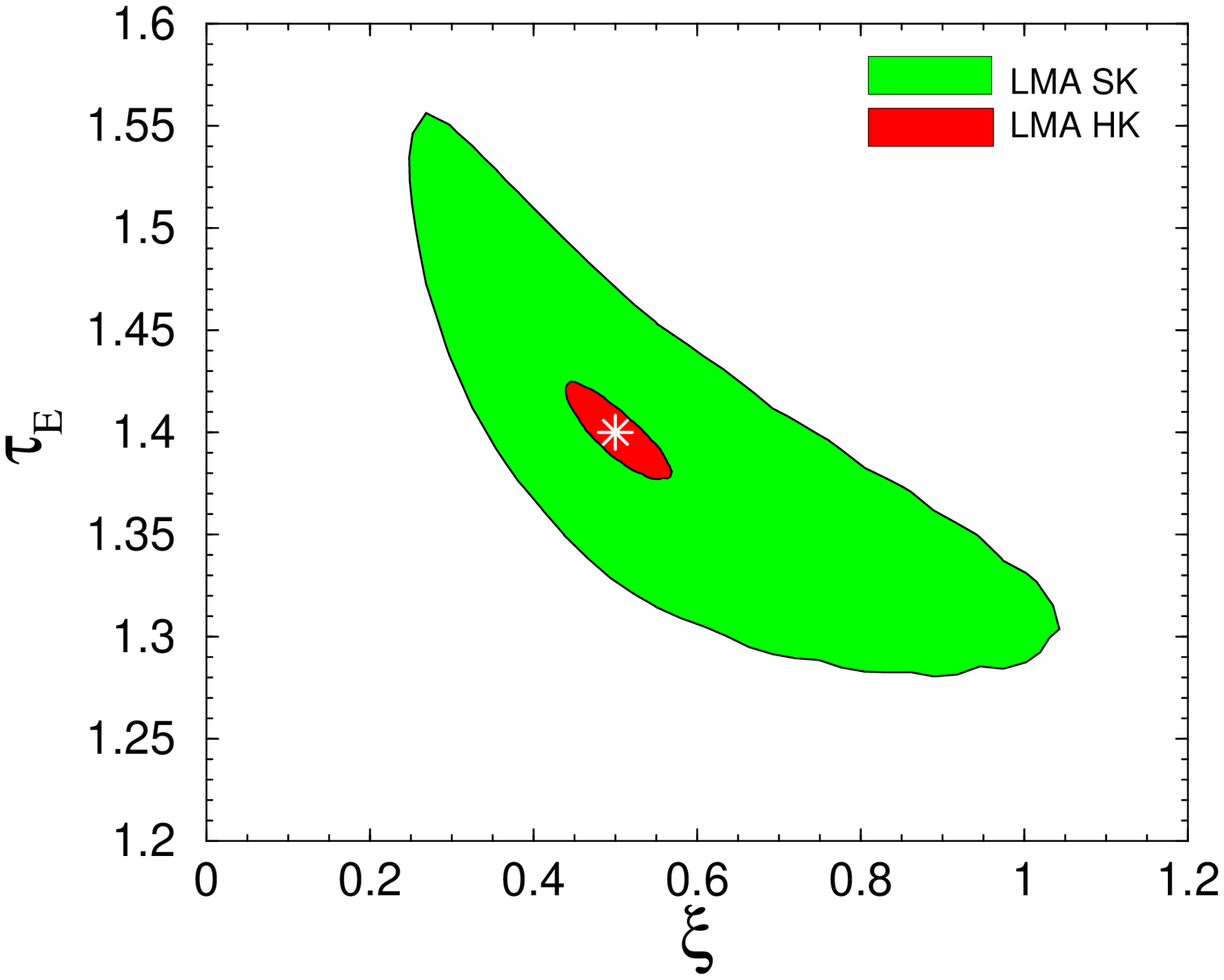}
\includegraphics[width=0.23\textwidth]{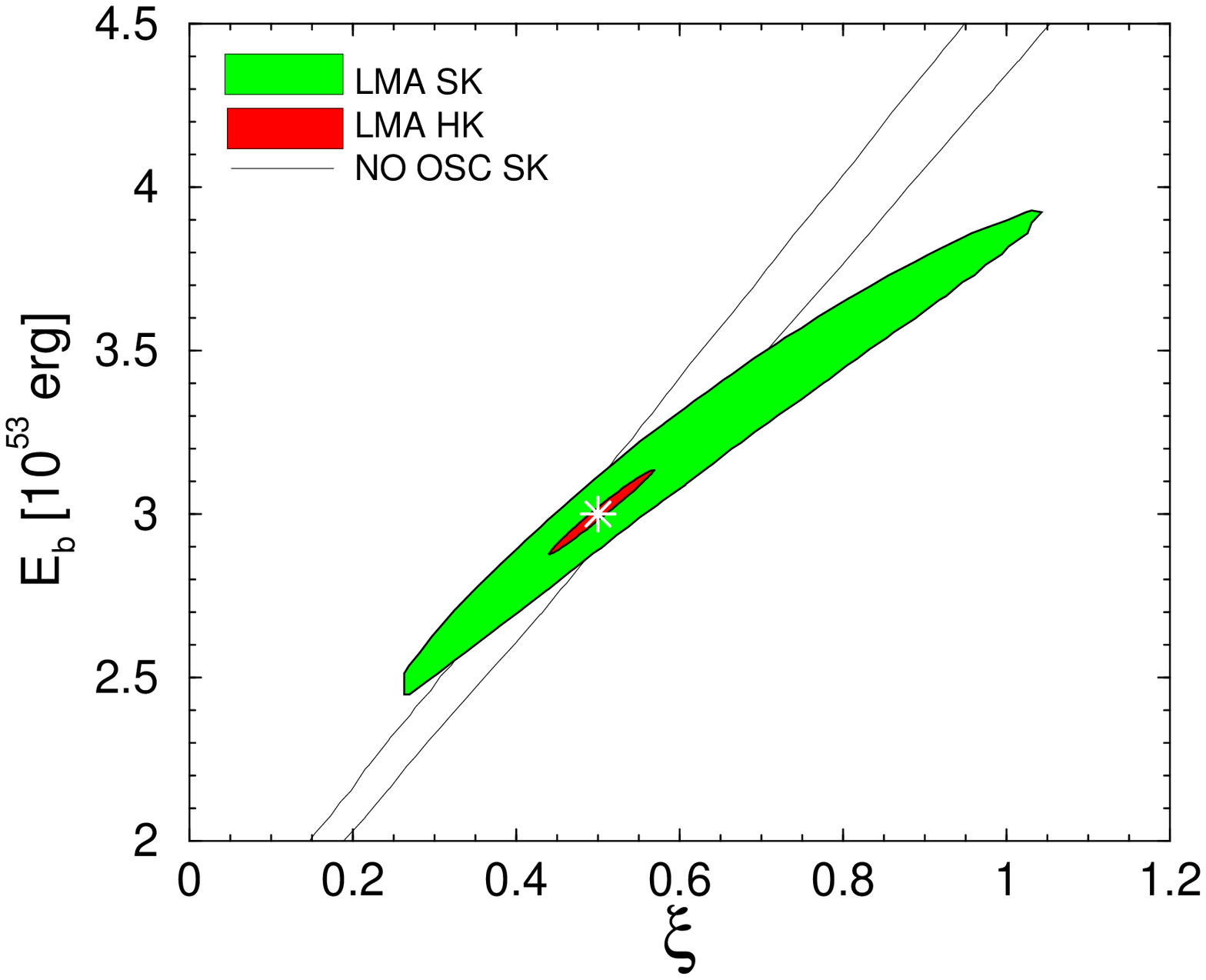}
\vspace*{-8mm}
\caption{Extracting supernova parameters from LMA oscillations~\cite{Minakata:2001cd}}
\label{minak}
\vspace*{-2mm}
\end{figure}

The large solar mixing indicated by present data will lead to
significant deformation of the energy spectra of supernova neutrinos,
affecting the resulting signal~\cite{Smirnov:ku}. However, a global
analysis shows that the LMA--MSW may remain as the best solution even
after combining SN1987 with solar neutrino
data~\cite{Kachelriess:2001sg}.
Finally we note that solar neutrino oscillations with large mixing may
allow us in the future to obtain otherwise inaccessible features of SN
neutrino spectra.  Fig.~\ref{minak} from~\cite{Minakata:2001cd} shows
how one can determine temperatures and luminosities of non-electron
flavor neutrinos by observing $\bne$ from a galactic supernova in
massive water Cherenkov detectors using the CC reactions on protons,
especially at a Hyper-K--type detector.

%%%%%%%%%%%%%%%%%%%%%%%%%%%%%%%%%%%%%%%%%%%%%%%%%%%%%%%%%%%%%%%%%%%%%%

\begin{figure}[t] 
    \includegraphics[width=0.46\textwidth,height=4cm]{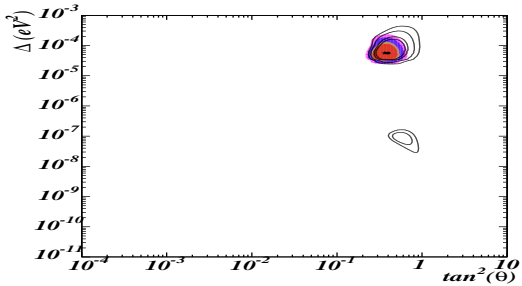}
\vspace*{-6mm}
    \caption{LMA as best solar + SN-1987A fit~\cite{Kachelriess:2001sg} }
\label{sn87sno02n.eps}
\vspace*{-5mm}
\end{figure}

%%%%%%%%%%%%%%%%%%%%%%%%%%%%%%%%%%%%%%%%%%%%%%%%%%%%%%%%%%%%%%%%%%%%%%

%%%%%%%%%%%%%%%%%%%%%%%%%%%%%%%%%%%%%%%%%%%%%%%%%%%%%%%%%%%%%%%%%%%%%%

\begin{figure}[htbp]
\includegraphics[width=0.4\textwidth,height=3.5cm]{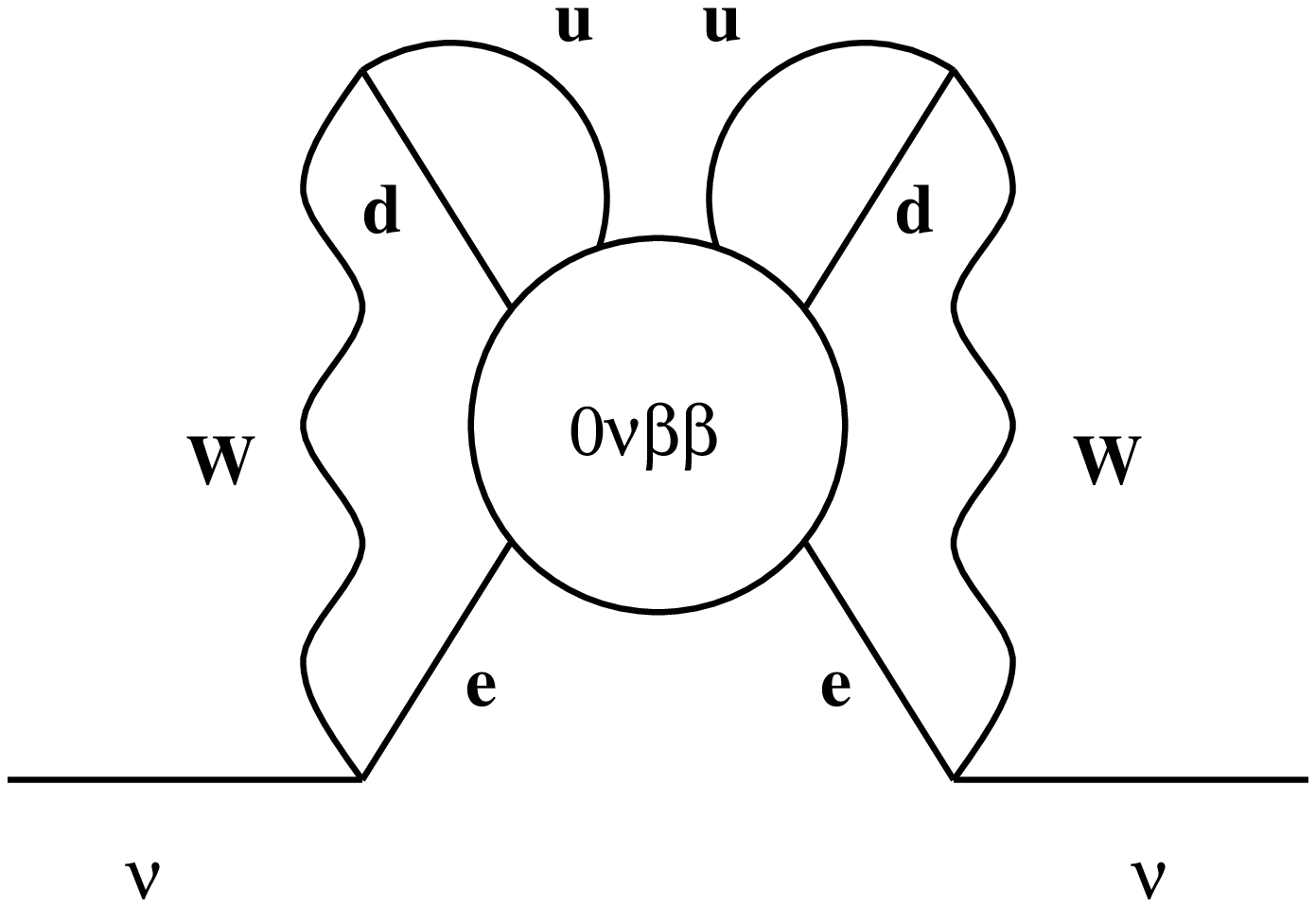}  
 \vspace*{-8mm}
 \caption{The black-box \nbb argument \cite{Schechter:1981bd}.}
 \label{fig:bbox} 
\vspace*{-8mm}
\end{figure}

\begin{figure}[htbp]
\includegraphics[width=0.4\textwidth,height=3.5cm]{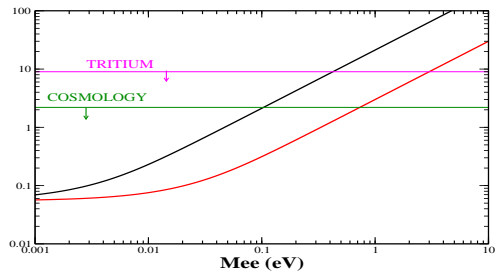}
 \vspace*{-8mm}
 \caption{\nbb and the scale of neutrino mass.}
 \label{fig:bbmass} 
 \vspace*{-6mm}
\end{figure}

Last, but not least, note that neutrino oscillations are sensitive
only to mass splittings, not to the absolute scale of neutrino mass,
nor to whether neutrinos are Dirac or Majorana particles.  The main
process relevant to decide this fundamental issue is \nbb decay
\cite{Morales:1998hu}.  The \texttt{black-box theorem} states that, in
a ``natural'' gauge theory, irrespective of how \nbb is engendered, it
implies a Majorana neutrino mass, and vice-versa, as illustrated by
Fig.  \ref{fig:bbox}.
One may quantify the implications of the black-box argument once a
particular model is specified. The strength of
neutrino-exchange-induced \nbb is characterized by an ``effective''
neutrino mass parameter $M_{ee}$ which takes into account possible
cancellations among individual neutrino
amplitudes~\cite{pseudo,Valle:1982yw}.  As seen in
Fig.~\ref{fig:bbmass} this is directly correlated~\cite{Barger:2002xm}
with the neutrino mass scales probed in tritium beta
decays~\cite{katrin} and cosmology~\cite{Elgaroy:2002bi}. It is
therefore important to probe \nbb in a more sensitive experiment, such
as GENIUS \cite{Klapdor-Kleingrothaus:1999hk}.
Moreover $\Delta L=2$ processes, such as \nbb decay, are sensitive to
the ``Majorana-type'' phases~\cite{Schechter:1980gr,Doi:1985dx} which
drop out from ordinary ($\Delta L=0$) neutrino oscillations. However
it is unlikely that these phases can ever be reliably extracted from
\nbb alone.
The CP violation induced by the ``Dirac'' phase is very hard to probe
in oscillations, since it disappears as two neutrinos become
degenerate \cite{Schechter:1979bn} and as $\theta_{13} \to
0$~\cite{CHOOZ}.  Fortunately the LMA solution helps, so with good
luck, neutrino factories may probe leptonic CP violating effects,
through the measurement of CP asymmetries~\cite{Lindner}.

%%%%%%%%%%%%%%%%%%%%%%%%%%%%%%%%%%%%%%%%%%%%%%%%%%%%%%%%%%%%%%%%%%%%%%

\section{NEUTRINO THEORIES}

Basic uncertainties hinder the prediction of neutrino masses from
first principles~\cite{Valle:1990pk}: no knowledge of the underlying
scale (the scale of gravity/strings? the GUT scale? an intermediate
left-right scale? the weak interaction scale itself?), no knowledge of
the underlying mechanism (tree level? radiative? hybrid?)  and, last
but not least, lack of a theory of flavor, which makes it especially
difficult to make an honest prediction of mixing angles.  Nevertheless
there has been an explosion of models in the last few years, most of
which based on the so-called seesaw scheme~\cite{seesaw,seesaw2}.  

An interesting of feature of such seesaw models is that the amount of
lepton asymmetry produced in the early universe by the
out--of--thermal--equilibrium decay of the right-handed neutrino may
be enough to explain the current baryon-to-photon ratio of the
Universe~\cite{Yanagida}. This asymmetry arises from leptonic CP
violation associated to the Majorana nature of
neutrinos~\cite{Schechter:1980gr}.

Various Yukawa textures and gauge groups have been proposed in order
to ``predict'' neutrino mixing angles in the seesaw approach, as
described by King~\cite{King}.  Here I simply mention a few models
from our own crop.  They fall into two classes: \texttt{top-down} and
\texttt{bottom-up}, and cover both \texttt{hierarchical} and
\texttt{quasi-degenerate} neutrino mass spectra.

\begin{figure}[t] \centering 
 \includegraphics[width=0.4\textwidth,height=3.5cm]{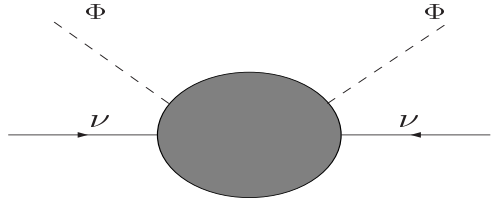}
\vspace*{-5mm}
    \caption{Dimension-5  neutrino mass operator}
\label{d-5.ps}
\vspace*{-3mm}
\end{figure}

\begin{figure}[t] 
\centering
    \includegraphics[width=0.4\textwidth,height=3.5cm]{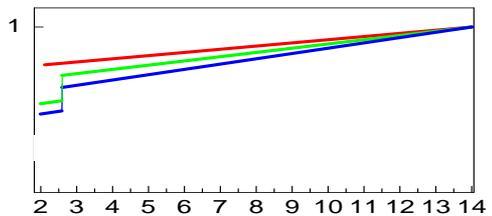}
\vspace*{-8mm}
    \caption{Neutrino masses unifying at $10^{14}$ GeV}
\label{neunif2.ps}
\vspace*{-5mm}
\end{figure}

The simplest way to give neutrino masses makes use of the basic
dimension-5 operator in Fig.~\ref{d-5.ps}, first noted by
Weinberg~\cite{Weinberg:uk} whose coefficient is unknown.  Since
gravity is believed to break global symmetries, such as lepton number,
it may induce this operator. Alternatively it may arise from physics
at some Grand-unified or intermediate scale, \texttt{a la seesaw}. 
A radical idea~\cite{Chankowski:2001fp} is that, due to some symmetry,
valid at some high-energy scale $M_X$, neutrino masses ``unify'' at
that scale, as indicated in Fig.~\ref{neunif2.ps}.
Such a simple ansatz naturally leads to quasi-degenerate neutrinos at
low scales, which could lie in the electron-volt range, while the tiny
solar and atmospheric neutrino mass splittings are induced by
renormalization effects.
An interesting choice for the underlying symmetry is the discrete
non-Abelian symmetry $A_4$~\cite{Babu:2002dz}, which can be realized
without spoiling the hierarchy of charged-lepton masses.
Solar and atmospheric neutrino mass splittings, maximal atmospheric
and large solar mixings are then induced radiatively in softly broken
supersymmetry. The quark mixing matrix is also calculable in a similar
way. The mixing parameter $U_{e3}$ is predicted to be small and
imaginary, leading to maximal CP violation in neutrino oscillations.
The \ne is a Majorana neutrino while the muon and tau neutrinos form a
pseudo-Dirac pair~\cite{pseudo}.  We found that \nbb and $\tau \to \mu
\gamma$ decay rates fall in the experimentally accessible range.

%%%%%%%%%%%%%%%%%%%%%%%%%%%%%%%%%%%%%%%%%%%%%%%%%%%%%%%%%%%%%%%%%%%%%%

Another minimalistic way to generate neutrino mass and mixings has
been suggested in \cite{deGouvea:2000jp}. It consists in producing the
atmospheric neutrino scale \texttt{a la seesaw} with just one \321
singlet lepton~\cite{Schechter:1980gr,Schechter:1979bn}.  In this
approximation two of the neutrinos are massless, their degeneracy
being lifted by the gravitationally induced dimension-5 operator
discussed above.  The required neutrino parameters can be easily
accommodated. Since the solar scale comes from Planck-mass effects,
the solar neutrino problem is explained by vacuum oscillations,
which may be tested through the search for anomalous seasonal effects
at Borexino.

%%%%%%%%%%%%%%%%%%%%%%%%%%%%%%%%%%%%%%%%%%%%%%%%%%%%%%%%%%%%%%%%%%%%%%

I now turn to the possibility that neutrino masses may have an
intrinsically supersymmetric origin, through the breaking of
R--parity~\cite{rpmnu-spo,rpmnu-exp,rpmnu-spopostLEP}.  I focus on the
bilinear violation of R--parity~\cite{Valle:1998be,Diaz:1997xc},
described by
\begin{equation}
  \label{eq:bilsup}
  W= W_{\rm MSSM} + \epsilon_a \ell_a H_u 
\end{equation}
where $ W_{\rm MSSM}$ is the MSSM superpotential, $\epsilon_a$ ($a=e,
\mu, \tau$) denote the strength of bilinear terms involving the lepton
($\ell_a$) and up-type Higgs ($H_{u}$) superfields.
The bilinear terms lead to \texttt{one} tree level neutrino mass,
chosen to lie in the range required by the atmospheric neutrino data,
while calculable radiative contributions lift the degeneracy of the
other neutrinos giving rise to the solar neutrino
scale~\cite{Hirsch:2000ef}.

This provides the simplest, most predictive and systematic effective
R--parity violation model at low--energies. Its theoretical basis can
be found either in the context of models where R--parity breaking is
introduced explicitly \texttt{ab initio}~\cite{rpmnu-exp}, or in
models where the violation of R--parity occurs \texttt{spontaneously},
through a non-zero \21 singlet sneutrino vacuum expectation
value~\cite{rpmnu-spopostLEP}.

A recent example of the former kind was given in~\cite{Mira:2000gg}
using an anomalous U(1) horizontal symmetry which forbids all
trilinear R-parity violating superpotential terms, selecting only the
bilinear ones, and relating their strength $\epsilon_a$ to powers of
the U(1) breaking parameter $\theta \sim 0.22$. This gives a common
origin for the $\mu$--term related to electroweak breaking and for the
L-violating terms generating neutrino masses.  The latter are suitable
for explaining neutrino anomalies, though radiative contributions
prefer the presently disfavored QVO solutions to the solar neutrino
problem. This can be tested through the search for anomalous seasonal
effects.  The neutrino mixing angles are not suppressed by powers of
$\theta$ and can naturally be large.

%%%%%%%%%%%%%%%%%%%%%%%%%%%%%%%%%%%%%%%%%%%%%%%%%%%%%%%%%%%%%%%%%%%%%%

Spontaneous R--parity breaking models require the addition of \21
singlet superfields, e.~g.~ right-handed neutrinos, and give a
dynamical origin for the bilinear strength $\epsilon_a$ identified as
$\epsilon_a = {h_\nu}_{a b} v_{Rb}$ where ${h_\nu}_{a b}$ is the Dirac
Yukawa matrix.  A characteristic feature of these models is the
existence of an novel variant (L=1) of the singlet seesaw
majoron~\cite{seesaw3}.  Majoron emission induces ``invisible''
neutrino~\cite{seesaw3} and Higgs boson decays~\cite{Romao:1992zx}.
The former are relevant in astrophysical
environments~\cite{Kachelriess:2000qc}, while the latter lead to
sizable Higgs--to--missing--transverse--momentum
signals~\cite{Carena:1996bj}.

\begin{figure}[t] 
\includegraphics[width=0.22\textwidth,height=4cm]{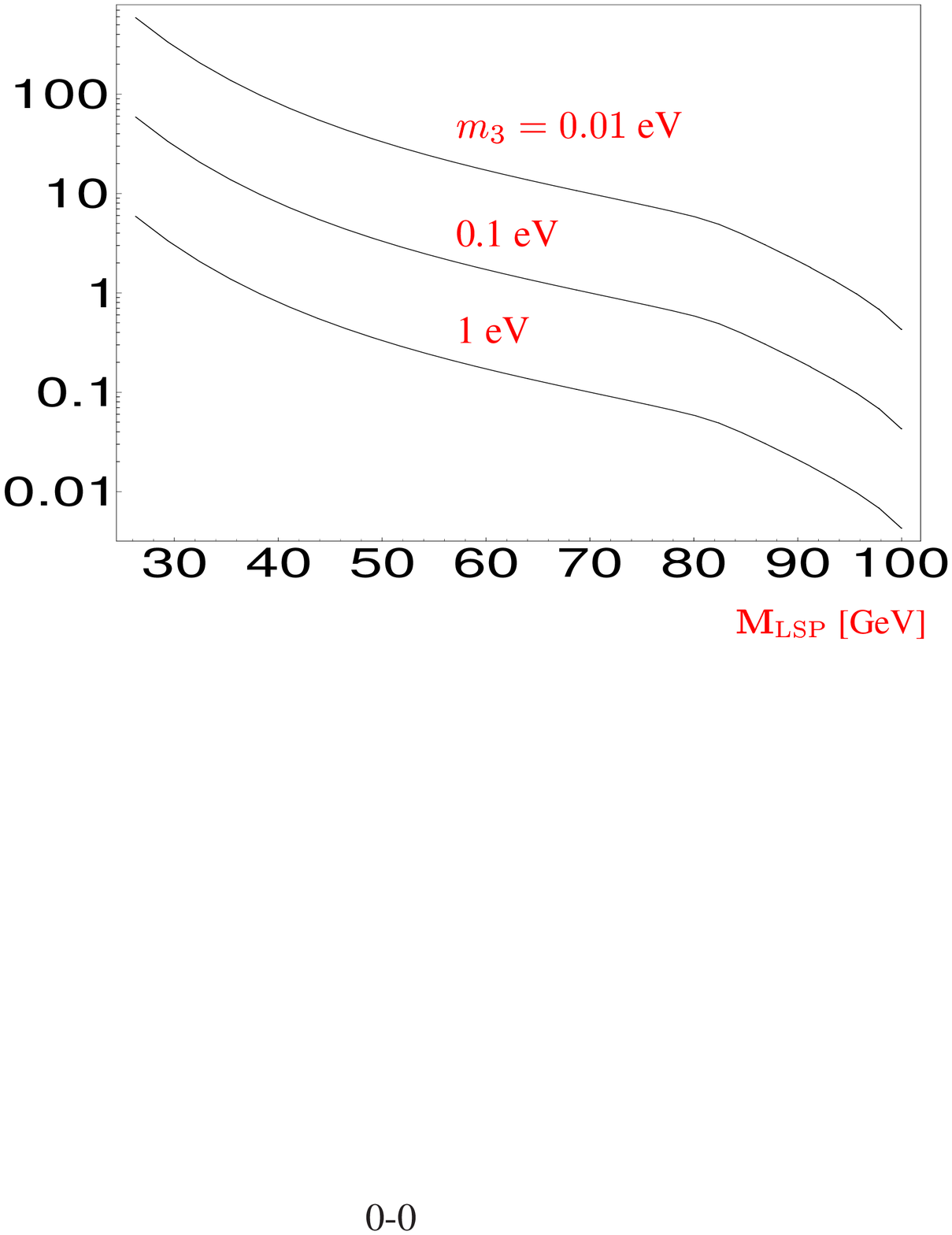}
\includegraphics[width=0.22\textwidth,height=4cm]{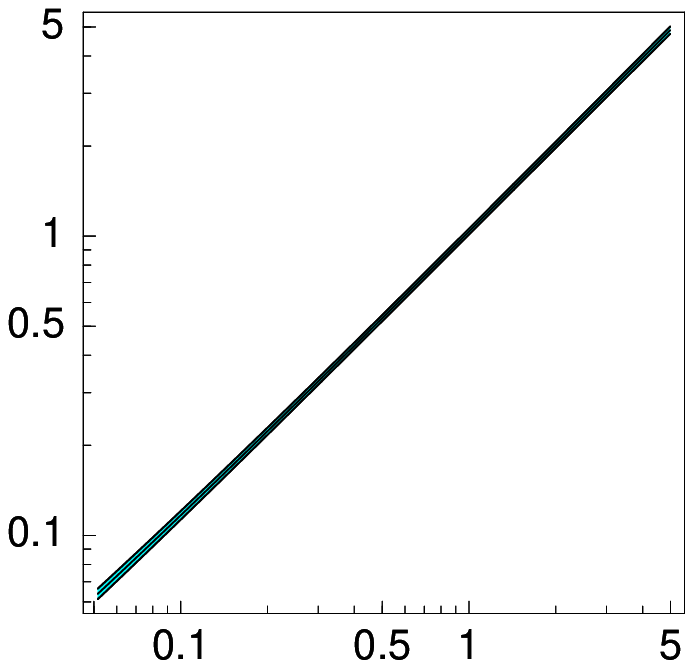}
\vspace*{-5mm}
\caption{LSP decay length~\cite{Bartl:2000yh} and predicted
  semileptonic decay branching ratios~\cite{Porod:2000hv}.}
\label{anomalytest}
\vspace*{-4mm}
\end{figure}

In addition to explaining neutrino anomalies, supersymmetry with
bilinear breaking of R--parity leads to a variety of phenomenological
implications~\cite{Allanach:1999bf}. Some examples are: unification
predictions for gauge and Yukawa couplings, $m_{\rm top}$, $V_{cb}$
and $\tan\beta$~\cite{Diaz:1999wm,Diaz:1999is,Diaz:1998wz}, $b \to s
\gamma$~\cite{Diaz:1998wq}, \nbb~\cite{Hirsch:2000jt}, charged and
neutral Higgs boson decays~\cite{Akeroyd:1997iq}, top quark
decays~\cite{Navarro:1999tz}, chargino~\cite{deCampos:1997mf} and
neutralino decays~\cite{Bartl:2000yh}, and gluino cascade
decays~\cite{Bartl:1996cg}. Barring fine-tuning or other assumptions,
the smallness of the neutrino masses suppresses many of these effects
in the most generic bilinear model of neutrino anomalies. However,
there is at least one which survives and which may lead to a dramatic
confirmation of the neutrino anomalies at high energy accelerator
experiments such as the LHC or NLC: the decays of the lightest
supersymmetric particle (LSP). This is a very striking feature of the
model which holds irrespective of what is the nature of the LSP. We
have considered the cases of lightest neutralino as
LSP~\cite{Porod:2000hv}, lightest stop as LSP ~\cite{Restrepo:2001me}
and lightest stau as LSP ~\cite{Hirsch:2002ys}.  As an example, the
left panel in Fig.~\ref{anomalytest} shows the LSP decay length in cm
versus mass in GeV, when it is the lightest neutralino.
%%%%%%%%%%%%%%%%%%%%%%%%%%%%%%%%%%%%%%%%%%%%%%%%%%%%%%%%%%%%%%%%%%%%%%
%%%%%%%%%%%%%%%%%%%%%%%%%%%%%%%%%%%%%%%%%%%%%%%%%%%%%%%%%%%%%%%%%%%%%%
On the other hand the right panel gives the ratio of predicted LSP
decay semileptonic decay branching ratios to muons over taus
(ordinate) versus the atmospheric neutrino mixing angle (abscissa),
illustrating a perfect correlation: if $\theta_\Atm = \pi/4$ one
expects equal numbers of muons and taus in semileptonic neutralino
decays.  Similarly, the solar and reactor neutrino angles can be
probed in various LSP decay scenarios, see
\cite{Porod:2000hv,Restrepo:2001me,Hirsch:2002ys} for details.

\section{NON-STANDARD NEUTRINOS}

Non-standard neutrinos interactions (NSI) are expected in most
neutrino mass models~\cite{Schechter:1980gr,Valle:1990pk} and can be
of two types: flavour-changing (FC) and non-universal (NU). They may
arise from a nontrivial structure of CC and NC weak interactions
characterized a non-unitary lepton mixing matrix and a correspondingly
non-trivial NC matrix~\cite{Schechter:1980gr}.  Such
\texttt{gauge-induced} NSI may lead to flavor and CP violation, even
with degenerate massless neutrinos \cite{NSImodels2}. In radiative
models of neutrino mass \cite{Zee:1980ai} and supersymmetric models
with broken R parity \cite{rpmnu-spo,rpmnu-exp} FC-NSI can also be
\texttt{Yukawa-induced}, from the exchange of spinless bosons.
In supersymmetric unified models, they may be calculable
as renormalization effects~\cite{NSImodels3}.

\begin{figure}[htbp]
  \centering
\includegraphics[scale=.24]{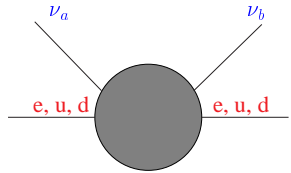}  
 \vspace*{-6mm}
 \caption{Effective  NSI operator.}
  \label{fig:nuNSI}
  \vspace*{-4mm}
\end{figure}

At the moment one can not yet pin down the exact profile of the \ne
survival probability and, as a result, the underlying mechanism of
solar neutrino conversion remains unknown.  Alternatives to
oscillations have been suggested since the eighties, including
non-standard neutrino matter interactions (NSI)~\cite{Valle:gv} and
spin-flavor precession (SFP)~\cite{Akhmedov:uk,Schechter:1981hw}. The
former may be represented as effective dimension-6 terms of the type
$\varepsilon G_F$, as illustrated in Fig.~\ref{fig:nuNSI}, where
$\varepsilon$ specifies their sub-weak strength.

\begin{figure}[t] 
\includegraphics[width=0.47\textwidth,height=3cm]{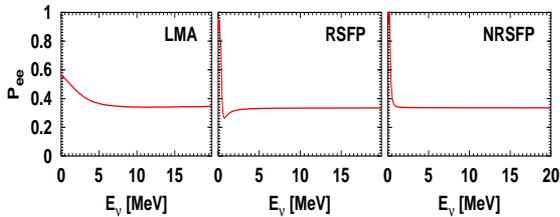}
\vspace*{-8mm}
\caption{Best LMA and SFP \ne survival probabilities from~\cite{Barranco:2002te}}
\label{probs.eps}
\vspace*{-4mm}
\end{figure}

Analyses of solar neutrino data in terms of NSI~\cite{Guzzo:2001mi}
and SFP~\cite{Barranco:2002te,Miranda:2001hv,Miranda:2000bi} have been
given recently.  SFP solutions exist both in the resonant
(RSFP)~\cite{Miranda:2000bi} and non-resonant regimes
(NRSFP)~\cite{Miranda:2001hv}.  The NSI solar neutrino energy spectrum
is undistorted, in agreement with data. On the other hand
Fig.~\ref{probs.eps} shows the predicted modification of the solar
neutrino spectra for the ``best'' LMA solution, and for the ``best''
SFP solutions \cite{Barranco:2002te}, from latest solar data. Clearly
the spectra in the high energy region are nearly undistorted in all 3
cases, in agreement with observations.

%%%%%%%%%%%%%%%%%%%%%%%%%%%%%%%%%%%%%%%%%%%%%%%%%%%%%%%%%%%%%%%%%%%%%%
\begin{figure}[t] 
\includegraphics[width=0.47\textwidth]{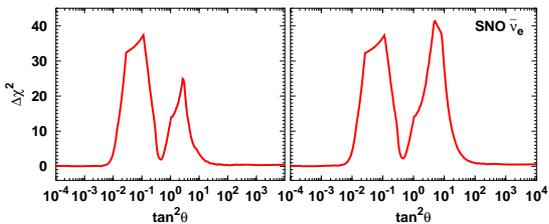}
\vspace*{-1cm}
\caption{$\Delta\chi^2_\Sol$ versus $\tan^2\theta_\Sol$, 
  for RSFP, LMA (central minima) and NRSFP solutions.  Left and right
  panels refer to two different analyses described in
  Ref.~\cite{Barranco:2002te}.}
\label{chi2sfp}
\vspace*{-6mm}
\end{figure}

\begin{figure}[h] 
\includegraphics[width=0.23\textwidth]{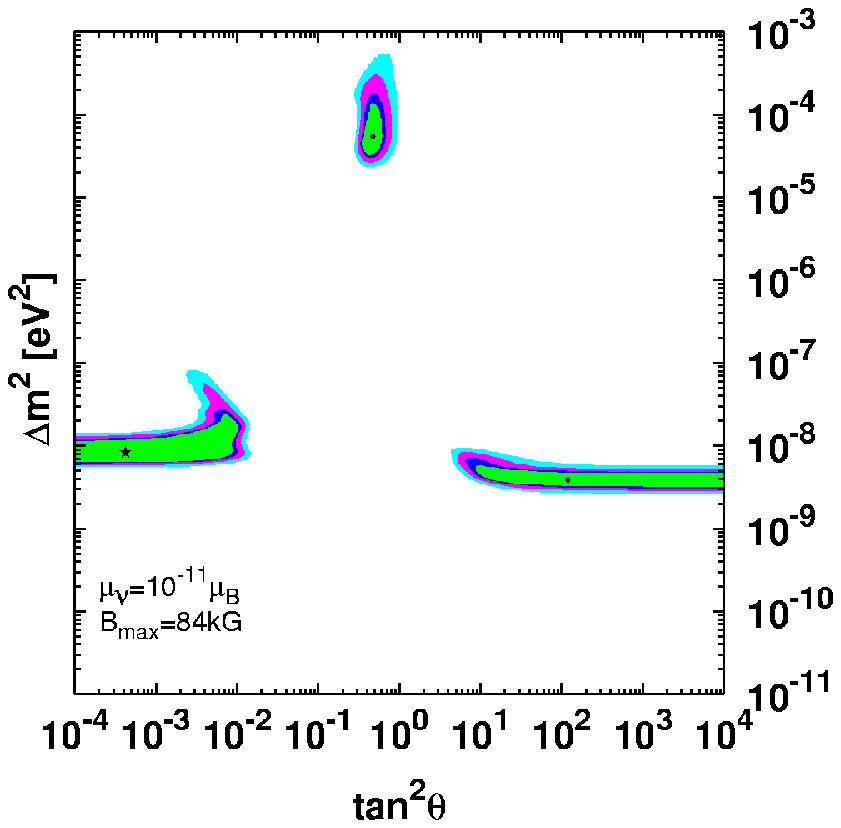}
\includegraphics[width=0.23\textwidth]{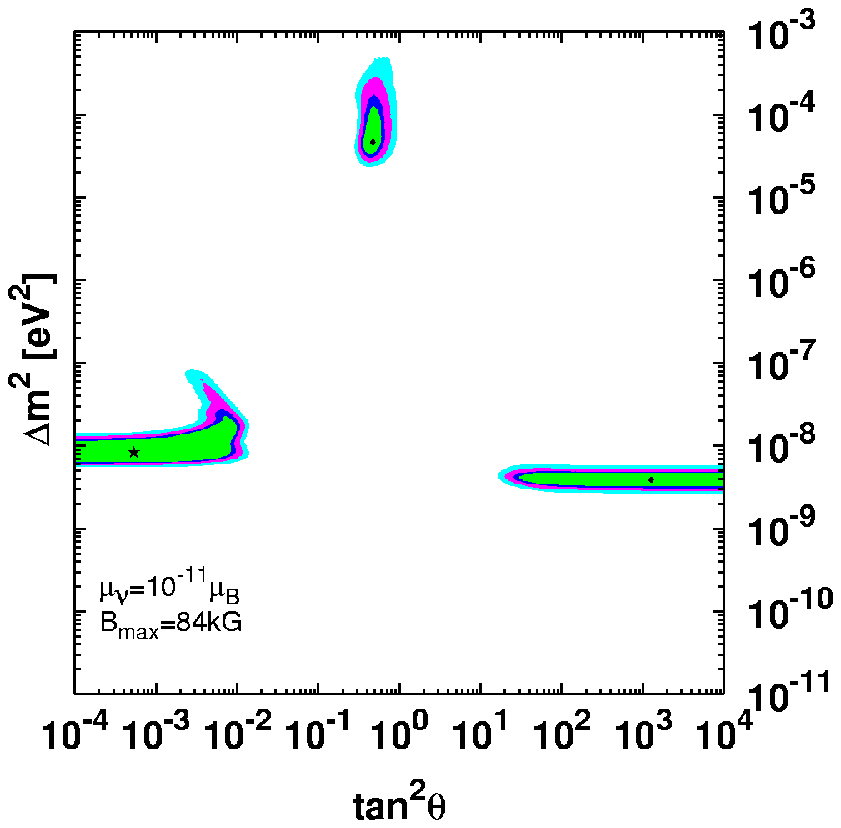}
\vspace*{-8mm}
\caption{Allowed $\Dms$ and $\tan^2\theta_\Sol$ for RSFP, LMA and NRSFP
  solutions for the indicated values of $\mu B$,
  from~\cite{Barranco:2002te}}
\label{sfpregions}
\vspace*{-2mm}
\end{figure}

%%%%%%%%%%%%%%%%%%%%%%%%%%%%%%%%%%%%%%%%%%%%%%%%%%%%%%%%%%%%%%%%%%%%%%
Although LMA oscillations are clearly favored over other
\texttt{oscillation} solutions~\cite{Maltoni:2002ni,Smirnov}, present
solar neutrino data \texttt{can be equally well--described by SFP and
  NSI solutions}. Fig.~\ref{chi2sfp} shows that this is indeed the
case, the resulting parameter regions is given in
Fig.~\ref{sfpregions}.
%%%%%%%%%%%%%%%%%%%%%%%%%%%%%%%%%%%%%%%%%%%%%%%%%%%%%%%%%%%%%%%%%%%%%%
%%%%%%%%%%%%%%%%%%%%%%%%%%%%%%%%%%%%%%%%%%%%%%%%%%%%%%%%%%%%%%%%%%%%%%
Although all 3 solutions are statistically equivalent, one sees that
that the two SFP solutions lie slightly lower than the LMA minimum.
%%%%%%%%%%%%%%%%%%%%%%%%%%%%%%%%%%%%%%%%%%%%%%%%%%%%%%%%%%%%%%%%%%%%%%
Note that in the presence of a neutrino transition magnetic moment of
$10^{-11}$ Bohr magneton, a magnetic field of 80 KGauss eliminates all
oscillation solutions other than LMA.
Ways to separate these 3 solutions at Borexino have been
considered in \cite{Barranco:2002te,Akhmedov:2002ti}.
%%%%%%%%%%%%%%%%%%%%%%%%%%%%%%%%%%%%%%%%%%%%%%%%%%%%%%%%%%%%%%%%%%%%%%

\begin{figure}[htbp]
  \centering
\includegraphics[width=0.47\textwidth]{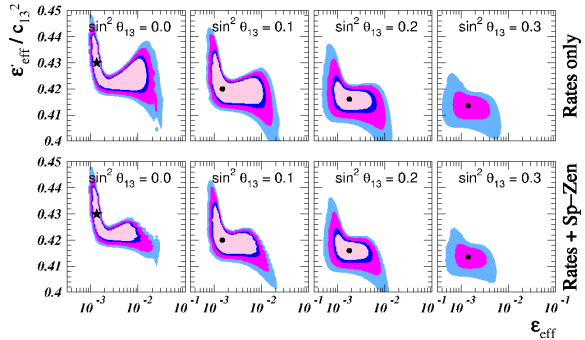}  
\vspace*{-8mm}
  \caption{Parameters of NSI solution to the solar neutrino anomaly, 
from \cite{Guzzo:2001mi}.}
  \label{fig:hybparam}
\vspace*{-6mm}
\end{figure}

Similarly the regions for the NSI mechanism are shown in
Fig.~\ref{fig:hybparam}. The required NSI values indicated by the
solar data analysis are fully acceptable also for the atmospheric
data.
Such NSI description of solar data was also shown~\cite{Guzzo:2001mi}
to be slightly better than that of the LMA oscillation solution.

%%%%%%%%%%%%%%%%%%%%%%%%%%%%%%%%%%%%%%%%%%%%%%%%%%%%%%%%%%%%%%%%%%%%%%

\begin{figure}[htbp]
\centering
\includegraphics[width=0.47\textwidth,height=5cm]{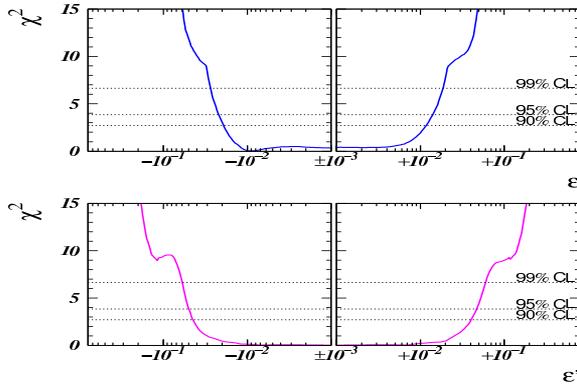}  
\vspace*{-8mm}
  \caption{Atmospheric bounds on NSI~\cite{Fornengo:2001pm}.}
  \label{fig:atmnsibds}
\vspace*{-5mm}
\end{figure}

Can NSI play a role in the atmospheric neutrino signal?  FC-NSI
interactions in the \nm-\nt channel without neutrino mass nor mixing
have been shown to account for the zenith--angle--dependent deficit of
atmospheric neutrinos observed in \texttt{contained} Super-K events
\cite{Fornengo:1999zp,Gonzalez-Garcia:1998hj}.  However such NSI
explanation fails to reconcile these with Super-K and MACRO
\texttt{up-going muons}, due to the lack of energy dependence
intrinsic of NSI conversions. As a result, a pure NSI conversion in
the atmospheric channel is ruled out at 99\% \CL
~\cite{Fornengo:2001pm}. Thus, unlike the case of solar neutrinos, the
oscillation interpretation of atmospheric data is robust, NSI being
allowed only at a sub-leading level.
Such robustness of the atmospheric \nm $\to$ \nt oscillation
hypothesis can be used to provide the most stringent present limits on
FC and NU neutrino interactions, as illustrated in
Fig.~\ref{fig:atmnsibds}.
These limits are also the most model--independent, as they are
obtained from pure neutrino-physics processes.

%%%%%%%%%%%%%%%%%%%%%%%%%%%%%%%%%%%%%%%%%%%%%%%%%%%%%%%%%%%%%%%%%%%%%%

Future neutrino factories aim at probing the lepton mixing angle
$\theta_{13}$ with much better sensitivity than possible at
present~\cite{Lindner}. They may also probe NSI in the \nm-\nt channel
\cite{Huber:2001zw}, with substantially improved sensitivity in the
case of FC-NSI, especially at energies higher than approximately 50
GeV.  For example, a 100 GeV \texttt{Nufact} can probe FC-NSI
interactions at the level of  $|\epsilon| < {\rm few} \times
10^{-4}$ at 99 \% C.L.

Note also that in such hybrid solution to the neutrino anomalies, with
FC-NSI explaining the solar data, and oscillations accounting for the
atmospheric data, the two sectors are connected not only by the
neutrino mixing angle $\theta_{13}$, but also by the \ne-\nt FC-NSI
parameters. As a result NSI and oscillations may be confused, as shown
in~\cite{Huber:2001de}.  This implies that information on
$\theta_{13}$ can only be obtained if bounds on NSI are available.
Taking into account the existing bounds on FC interactions, one finds
a drastic loss in \texttt{Nufact} sensitivities on $\theta_{13}$, of
at least two orders of magnitude. A near--detector offers the
possibility to obtain stringent bounds on some NSI parameters and
therefore constitutes a crucial necessary step towards the
determination of $\theta_{13}$ and subsequent study of leptonic CP
violation.

%%%%%%%%%%%%%%%%%%%%%%%%%%%%%%%%%%%%%%%%%%%%%%%%%%%%%%%%%%%%%%%%%%%%%%

Last, but not least, note that the KamLAND experiment~\cite{KamLAND}
will provide vital information very soon.  Even if the LMA solution is
finally confirmed by KamLAND, such alternative mechanisms will still
be interesting for study since their sub-leading admixture may be
testable, as discussed in~\cite{Barranco:2002te,Grimus:2002vb}.  They
may also affect the propagation of neutrinos in a variety of
astrophysical environments~\cite{Nunokawa:1996tg}.

%%%%%%%%%%%%%%%%%%%%%%%%%%%%%%%%%%%%%%%%%%%%%%%%%%%%%%%%%%%%%%%%%%%%%%

\end{document}